\documentclass[aps,preprintnumbers,twocolumn,amsmath,amssymb,prl,floatfix,superscriptaddress]{revtex4-2}

\usepackage{graphicx}
\usepackage{dcolumn}
\usepackage{bm}
\usepackage{graphicx}
\usepackage{caption}
\usepackage{multirow}
\usepackage{subcaption}
\usepackage{ragged2e}
\usepackage{cancel}
\usepackage{comment}
\DeclareCaptionJustification{justified}{\justifying}
\usepackage[hidelinks]{hyperref}
\usepackage{amssymb}
\usepackage{amsmath}

\begin{document}

\title{The thermal bootstrap for the critical $\mathrm O(N)$ model}

\author{Julien Barrat}\email{Email: julien.barrat@desy.de}
\affiliation{  Deutsches Elektronen-Synchrotron DESY, Notkestr. 85, 22607 Hamburg, Germany
}
\author{Enrico Marchetto}\email{Email: enrico.marchetto@desy.de}
\affiliation{Mathematical Institute, University of Oxford,
	Andrew Wiles Building, Woodstock Road, Oxford, OX2 6GG, U.K.
}%
\affiliation{  Deutsches Elektronen-Synchrotron DESY, Notkestr. 85, 22607 Hamburg, Germany
}
\author{Alessio Miscioscia}\email{Email: alessio.miscioscia@desy.de}
\affiliation{
 Deutsches Elektronen-Synchrotron DESY, Notkestr. 85, 22607 Hamburg, Germany
}%
\author{Elli Pomoni}\email{Email: elli.pomoni@desy.de}
\affiliation{
 Deutsches Elektronen-Synchrotron DESY, Notkestr. 85, 22607 Hamburg, Germany
}%

\date{November 1, 2024}

\begin{abstract}
\noindent  We propose a numerical method to estimate one-point functions and the free-energy density of conformal field theories at finite temperature by solving the Kubo--Martin--Schwinger condition for the two-point functions of identical scalars. We apply the method for the critical $\mathrm O(N)$ model for $N = 1,2,3$ in $3 \leq d \leq 4$.
We find agreement with known results from Monte Carlo simulations and previous results for the $3d$ Ising model, and we provide new predictions for $N= 2,3$.
\end{abstract}

\preprint{DESY-24-167}

\maketitle

\paragraph{Introduction and summary -}
Finite-temperature phenomena in conformal field theories (CFTs) can be studied by placing the theory on the geometry $S_\beta^1 \times \mathbb R^{d-1}$, where $\beta = 1/T$ is the inverse temperature.
Thermal dynamics plays a crucial role, as quantum critical points in experimental systems occur at non-zero temperatures \cite{Sachdev:2011fcc,Vojta:2003tcv}.
Additionally, it is essential to study CFTs at finite temperature to gain insights on Anti-de Sitter black holes in the quantum regime \cite{Witten:1998zw}.

The success of the conformal bootstrap in constraining zero-temperature CFT data (see, e.g., the reviews \cite{Simmons-Duffin:2016gjk,Poland:2018epd,Rychkov:2023wsd}), namely conformal dimensions and structure constants, naturally raises the question of whether similar techniques can be applied to thermal CFTs \cite{El-Showk:2011yvt,Iliesiu:2018fao}.
Since the operator product expansion (OPE) of the original CFT remains valid locally 
\footnote{The OPE holds operatorially, though its radius of convergence is finite (and equal to $\beta$) at finite temperature.}, thermal correlation functions can be expressed in terms of zero-temperature CFT data and thermal one-point functions. 
The goal of the thermal bootstrap program is to compute these observables employing the zero-temperature data as an input, and the Kubo--Martin--Schwinger (KMS) condition \cite{Kubo:1957mj,Martin:1959jp}, namely the periodicity along the thermal circle, as a consistency constraint.
Among all the operators, a special role is played by the stress-energy tensor, since its thermal one-point function is closely related to the free-energy density of the system \cite{Iliesiu:2018fao,Barrat:2024aoa}.

In this letter, we introduce a new efficient 
method to numerically estimate thermal one-point functions.
We impose the KMS condition on a thermal two-point function of identical scalars near the KMS fixed point \footnote{The KMS fixed point is achieved when the two operators are placed at an imaginary time separation of $\tau = \beta/2$.}. This generates an infinite set of equations with an infinite number of unknowns.
The novelty of this work is to analytically approximate the contribution of heavy operators using an improved version of the Tauberian asymptotics proposed in \cite{Marchetto:2023xap}, reducing the system to a finite set of unknowns.

\begin{figure}[t]
 \includegraphics[width=86mm]{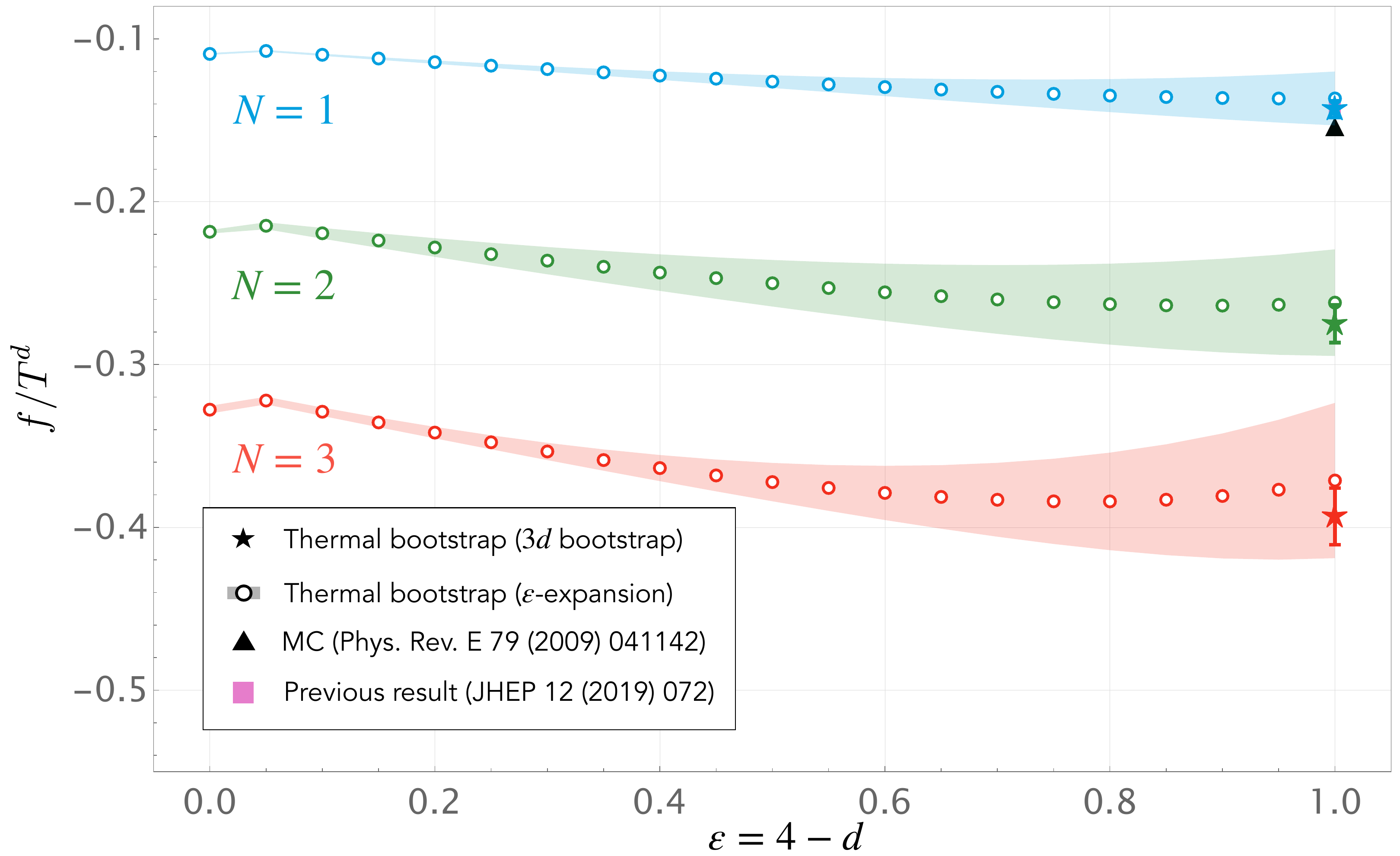}
    \caption{Free-energy density of the critical $\mathrm O(N)$ models for $N = 1,2,3$ in $3 \le d\le 4$ (i.e., $0 \le \varepsilon \le 1$).}
    \label{fig:Free_Energy}
\end{figure}

\begin{figure}[t]
 \includegraphics[width=85mm]{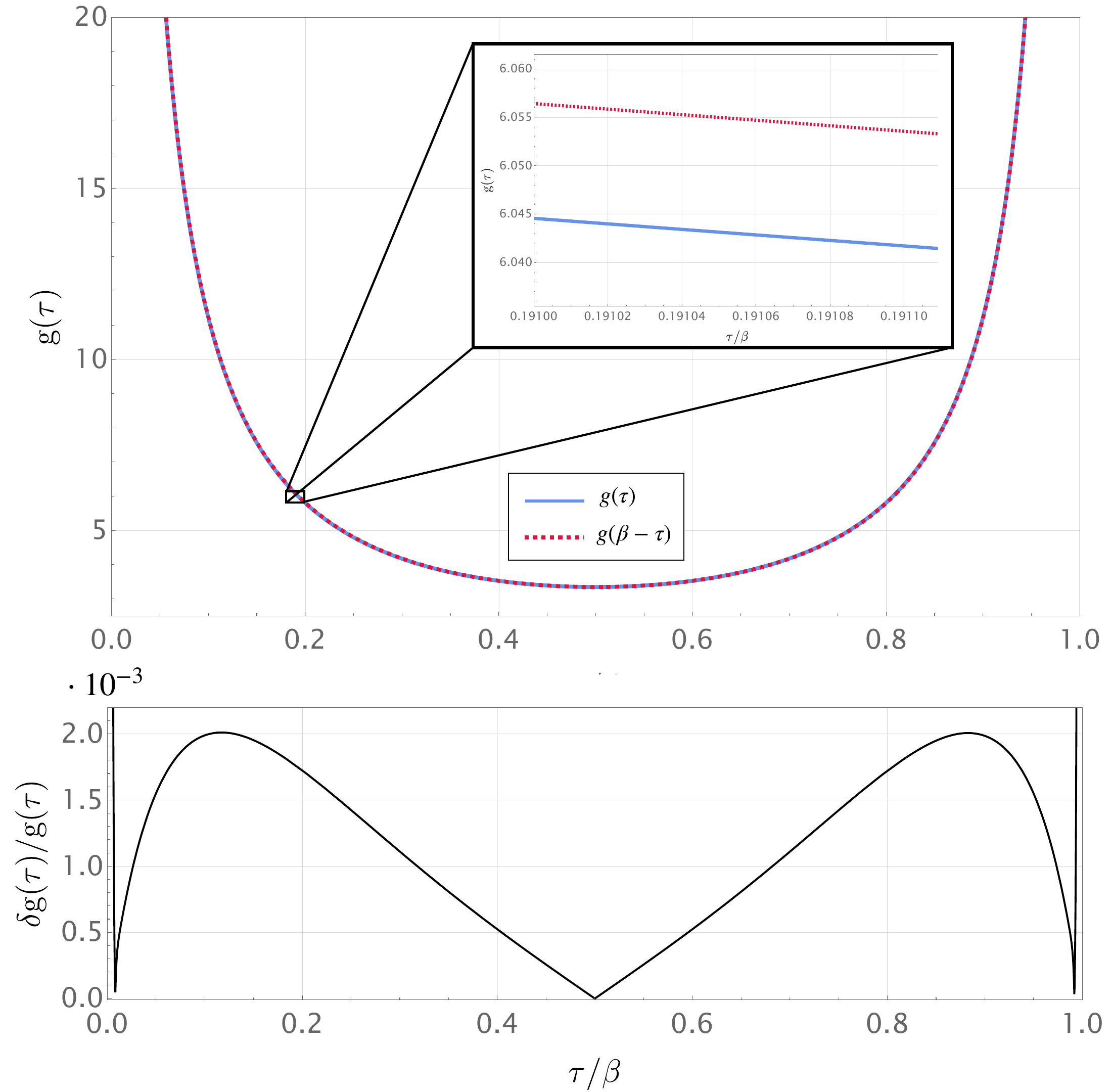}
    \caption{The thermal two-point function $g(\tau)$ is shown alongside its KMS equivalent $g(\beta - \tau)$ in the Ising model.
    The second plot shows the difference between the two curves in the main plot.
    We observe an excellent agreement in the region around the KMS fixed point ($\tau/\beta = 1/2$). }
    \label{fig:2_pt}
\end{figure}

\begin{figure}[t]
 \includegraphics[width=85mm]{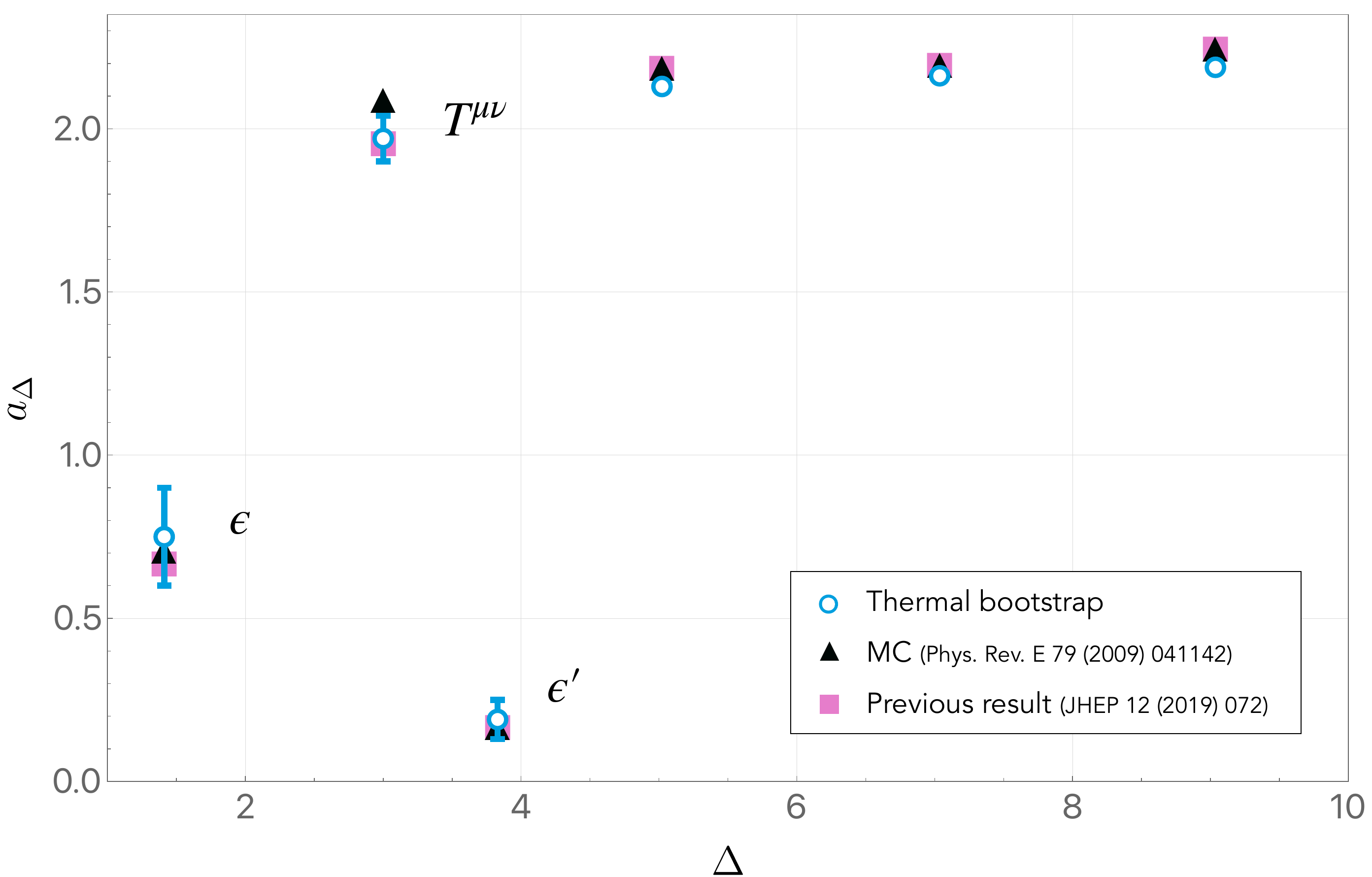}
    \caption{OPE coefficients for the lightest operators of the critical $3d$ Ising model. The points with no error bar associated correspond to analytical Tauberian predictions, whose error is not the object of study of this letter.}
    \label{fig:OPE_Ising}
\end{figure}

\begin{figure*}
    \begin{subfigure}[t]{.48\textwidth}
  \includegraphics[width=85mm]{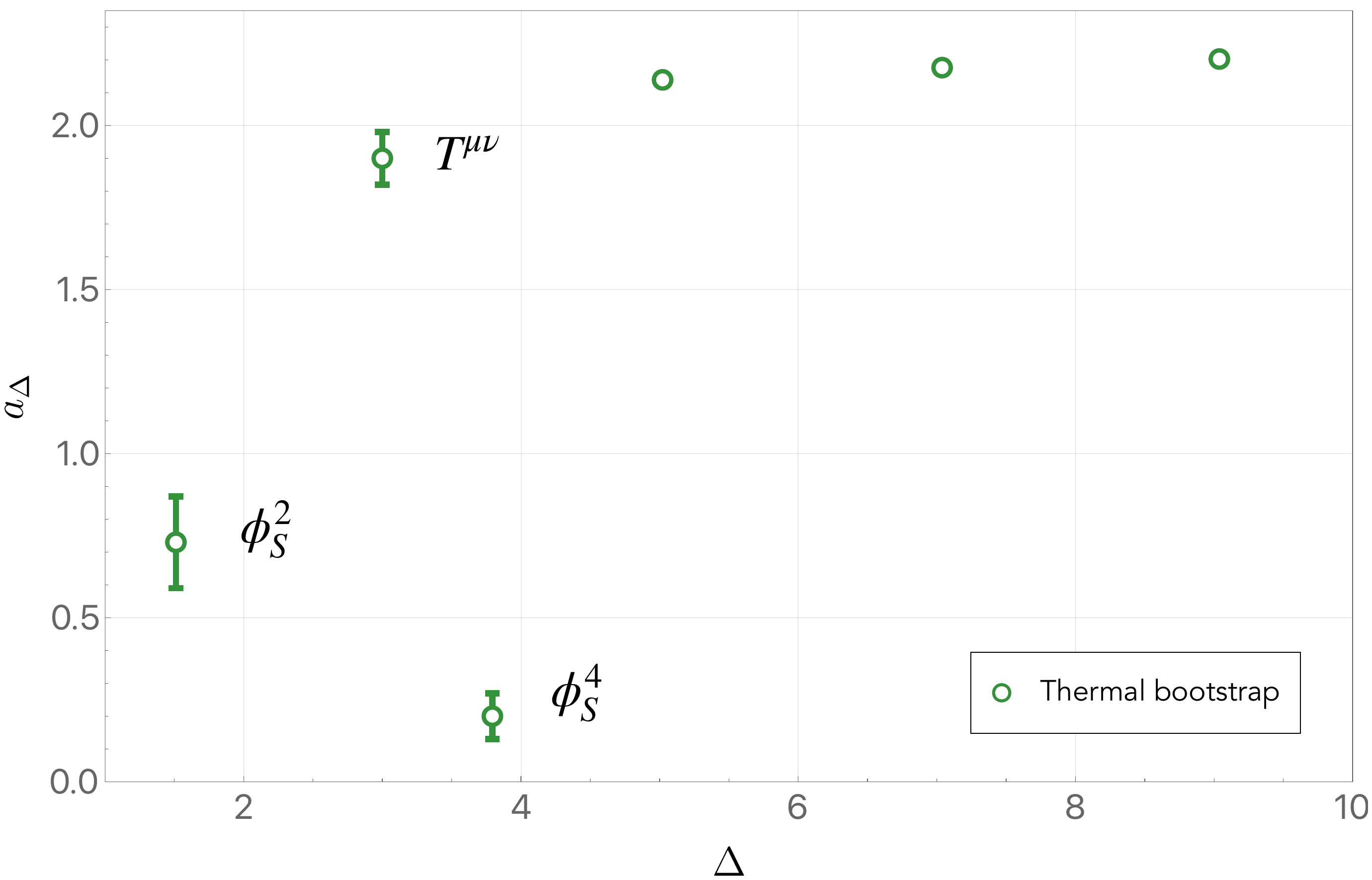}
  \caption{}
\end{subfigure}%
\hfill
\begin{subfigure}[t]{.48\textwidth}
  \includegraphics[width=85mm]{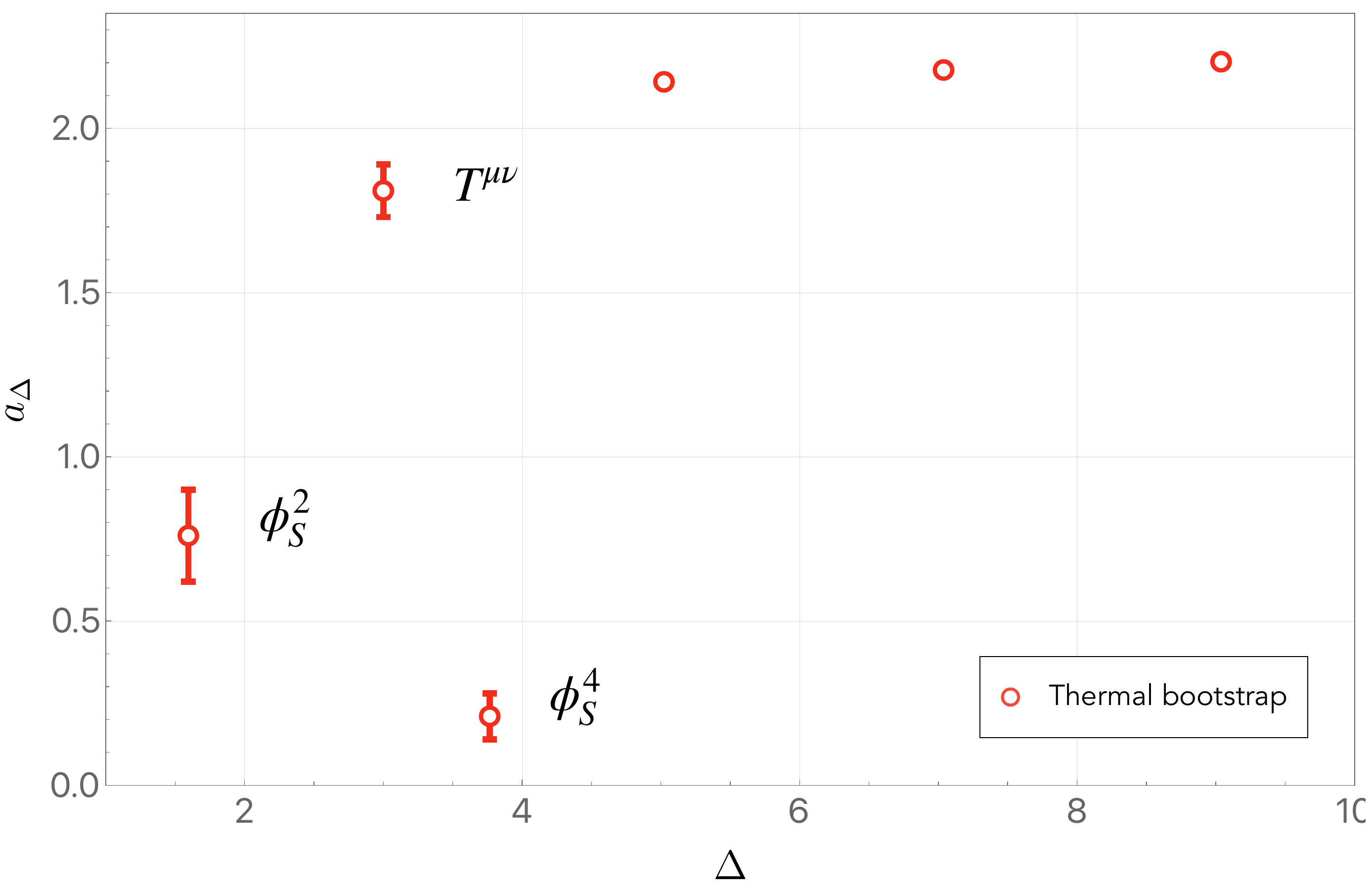}
  \caption{}
\end{subfigure}
    \caption{The two plots present the results for the OPE coefficients associated to the lightest operators of the OPE spectrum for the $\mathrm{O}(2)$ (left) and $\mathrm{O}(3)$ (right) models. The points with no error bar associated correspond to analytical Tauberian predictions, whose error is not the object of study of this letter.}
    \label{fig:OPE_Coefficients}
\end{figure*}

The method can be tested in $4d$ free scalar theory, $2d$ Ising model and in the large $N$ limit of the $\mathrm O(N)$ model, where numerical estimations can be compared with analytical results \cite{SM}. In the following we apply it in the strongly-coupled regimes of the critical $\mathrm{O}(N)$ models for $N=1, 2, 3$. These correspond to the critical Ising model ($N=1$), the XY model ($N=2$), and the Heisenberg model ($N=3$), which are relevant for understanding ferromagnetism and other physical phenomena  \cite{Berezinsky:1970fr,Kosterlitz:1973xp,McBryan:1976pk,PhysRevLett.77.940}.
Our key results are: the free-energy density in $3 \leq d \leq 4$ (Fig. \ref{fig:Free_Energy}), the two-point function of the lightest scalar in the critical $3d$ Ising model (Fig. \ref{fig:2_pt}), and the one-point functions of several operators in the critical $\mathrm O(1)$, $\mathrm O(2)$ and $\mathrm O(3)$ models (Figs. \ref{fig:OPE_Ising}, \ref{fig:OPE_Coefficients}).
In the case of the $3d$ Ising model, our results can be compared with previous numerical estimates \cite{Iliesiu:2018zlz} and Monte Carlo simulations \cite{PhysRevE.79.041142,PhysRevE.53.4414,PhysRevE.56.1642}, confirming the validity of our method \footnote{Thermal OPE coefficients from Monte Carlo simulations require combining simulation results with the inversion formula, as done for the $3d$ Ising model in \cite{Iliesiu:2018zlz}.}.
The predictions for $N=2, 3$ are new and could, in principle, be tested through further Monte Carlo simulations or experiments \cite{PhysRevB.90.184413}.

\bigskip

\paragraph{Thermal bootstrap -}
The starting point of our analysis is the KMS condition.
For the two-point function of identical scalar operators $g(\tau) = \langle \phi(\tau) \phi(0) \rangle_\beta$, where the spatial distance between the two operators is set to zero, the KMS condition results into a tower of constraints that take the form
\begin{equation}
   0= \frac{\partial^m}{\partial \tau^m} \left[ g \left( \frac{\beta}{2}+\tau \right)
    -
    g \left( \frac{\beta}{2}-\tau \right) \right]_{\tau = 0}
    \,,
    \label{eq:Derivatives}
\end{equation}
where $m \in 2 \mathbb N + 1$. 
These constraints can be expressed as a set of sum rules \cite{Marchetto:2023xap}
\begin{equation}
    \sum_{\Delta} a_\Delta \  \mathtt {F}(\Delta,\Delta_\phi,m)
    =
    0\,,
    \label{eq:SumRules}
\end{equation}
where the sum is performed over all the operators in the OPE between the two operators $\phi$.
The kernel $\mathtt{F}$, defined in Eq. (9) in \cite{Marchetto:2023xap}, depends solely on zero-temperature CFT data, which we treat as input.
Meanwhile, the coefficients $a_\Delta$ encode the thermal dynamical information
\begin{equation}
    a_\Delta
    =
    \sum_{\mathcal O_\Delta} \frac{b_{\mathcal O} f_{\phi \phi \mathcal O}}{c_{\mathcal O}} \frac{J! }{2^J(\nu)_J} C_J^\nu(1) \,,
    \label{eq:aDelta}
\end{equation}
where $\nu=(d-2)/2$, $C_J^{(\nu)}$ is a Gegenbauer polynomial and the sum is performed over operators sharing the same scaling dimension, but with different spins.
Here, the coefficients $f_{\phi \phi \mathcal O}$ and $c_{\mathcal O}$  correspond, respectively, to the structure constants and to the two-point function normalization of the operator $\mathcal O$ at zero temperature.  $b_{\mathcal O}$ is the thermal one-point function coefficient defined via \cite{Iliesiu:2018fao,Marchetto:2023fcw}
\begin{equation}
    \langle
    \mathcal{O}_{\Delta}^{\mu_1 \ldots \mu_J} 
    \rangle_\beta
    =
    \frac{b_{\mathcal{O}}}{\beta^{\Delta}} (e^{\mu_1} \ldots e^{\mu_J}
    -
    \text{traces})\,.
    \label{eq:OnePointFunctions}
\end{equation}
The ultimate goal of the thermal bootstrap program is to compute these observables completing the set of thermal CFT data.

In order to solve the constraints \eqref{eq:SumRules}, a naive approach consists in truncating the sum at a cut-off dimension $\Delta_\text{max}$.
However this approach fails, as the contribution of the heavy operators cannot be discarded \footnote{The error introduced by neglecting the tail of heavy operators is shown in \cite{Marchetto:2023xap}. This is very different from the zero-temperature scenario, since in such case the naive truncation of the crossing equations can still lead to reasonably good approximations \cite{Gliozzi:2013ysa,Gliozzi:2014jsa,Gliozzi:2016cmg}. Nevertheless, the tail of heavy operators is still important to achieve a higher precision, as shown by \cite{Su:2022xnj,Li:2023tic,Poland:2023bny}. In particular, in \cite{Li:2023tic,Poland:2023bny} the authors employ a procedure similar to ours for the estimation of the error.}.
This issue can be circumvented by approximating the tail of heavy operators using the asymptotic behavior of the coefficients $a_\Delta$ \cite{Marchetto:2023xap}
\begin{equation}
    a_\Delta ^{\text{heavy}}
    =
    \frac{\Delta^{2\Delta_\phi-1}}{\Gamma(2\Delta_\phi+1)}\delta \Delta
    \left(1+\frac{c_1}{\Delta}
    +
    \ldots \right)\,.
    \label{eq:Tauberian}
\end{equation}
Here, $\delta\Delta$ represents the gap between the scaling dimension $\Delta$ and the scaling dimension of the operator below it in the OPE spectrum.
The coefficient $c_1$ is theory-dependent and corresponds to the first correction to the leading behavior.
Let us comment that, in order to derive \eqref{eq:Tauberian}, it is necessary to add an analyticity assumption on $a_{\Delta}$, since the Tauberian theorem fixes only the leading term \cite{Marchetto:2023xap}.
Moreover, note that the power of $\Delta$ in the first correction is universal, but those of the sub-leading terms are theory-dependent and currently unknown. Determining them is an important next goal.
The constraints of Eq. \eqref{eq:SumRules} can be split into two terms:
\begin{equation}
    \mathtt f(m)
    =
    \sum_{\Delta \leq \Delta_\text{max}} \hspace{-0.75em} a_\Delta  \mathtt {F}(\Delta,\Delta_\phi,m)
    +
    \sum_{\Delta > \Delta_\text{max}} \hspace{-0.75em}a_\Delta^{\text{heavy}}  \mathtt {F}(\Delta,\Delta_\phi,m)\,.
    \label{eq:ConstraintsWithTauberian}
\end{equation}
We do not have access to the spectrum in the heavy sector and for this reason a further approximation is required.
In this paper, we restrict our consideration to the leading trajectories of operators $[\phi\phi]_{n,\ell}$ in the second term of Eq. \eqref{eq:ConstraintsWithTauberian}, which, by channel duality, correspond to the identity \footnote{The identity block inverts to the double-twist operators in both the zero- and finite-temperature Lorentzian inversion formulas \cite{Simmons-Duffin:2016wlq,Iliesiu:2018fao,Fitzpatrick:2012yx,vanRees:2024xkb}.}.These operators take the classical form $\phi \partial^\ell \Box^n \phi$, and their conformal dimensions can be approximated by the mean-field theory result $\Delta_{n,\ell} = 2 \Delta_\phi + 2n + \ell$. Thus, there are two sources of error: one arises from the omission of sub-leading trajectories, the other from the anomalous dimensions of these operators. The former is negligible with respect to the latter: we estimate both in \cite{SM}.
In this approximation, only a \textit{finite} number of unknown coefficients are left: the coefficients $a_\Delta$ associated with the light operators $\Delta \leq \Delta_\text{max}$, and the corrections to the leading behavior \eqref{eq:Tauberian}, namely $c_1$, $\ldots$ .
The constraints \eqref{eq:SumRules} can be formulated as the minimization of the cost function
\begin{equation}
    \eta(\{\mathtt \omega_i\})
    =
    \sum_{m \leq m_\text{max}}  \hspace{-0.60em} \mathtt \omega_m \mathtt f(m)^2\,,
    \label{eq:CostFunction}
\end{equation}
where $m_{\mathrm{max}}$ determines the maximum number of derivatives considered and $\omega_i \in (0,1)$ is a set of random number weights, which allows us to test the numerical stability of the algorithm as previously done, e.g., in \cite{Poland:2023bny}.
The minimization process results in estimations for the unknown parameters, which are affected by numerical errors stemming from two contributions:
\begin{itemize}
    \item A \textit{statistical} error, estimated by the square root of the variance over multiple runs of the minimization of \eqref{eq:CostFunction};
    \item A \textit{systematic} error, due to the approximation of the contribution of the heavy operators using \eqref{eq:Tauberian}, estimated in \cite{SM}. 
\end{itemize}
The uncertainties in the zero-temperature input data propagate into the thermal one-point functions. When using the $3d$ bootstrap results as input, these errors remain negligible. However, when the zero-temperature conformal data comes from the $\varepsilon$-expansion, the associated error increases with $\varepsilon$. This effect is estimated in \cite{SM} and illustrated in Fig. \ref{fig:Free_Energy} for $0 \leq \varepsilon \leq 1$.
The errors presented in this letter should be regarded as estimations rather than rigorous error bounds, adopting the terminology of \cite{Chang:2024whx}.

The free-energy density of the system is determined by the one-point function coefficient of the stress-energy tensor through $f = b_T/d$, with $d$ the number of spacetime dimensions \cite{Iliesiu:2018fao}.
The structure constant $f_{\phi \phi T}$, appearing in \eqref{eq:aDelta}, is fixed by the Ward identity \footnote{We assume here that the only operator with conformal dimension $\Delta = d$ is the stress-energy tensor as in the case of the $\mathrm{O}(N)$ model that we are studying in this letter in $2\leq d \leq 4$.} and therefore
\begin{equation}
    f
    =
    -
    a_d
    \frac{\Gamma\left(d/2\right)}{2 \pi^{d/2} (d-1)\Delta_\phi}
    \frac{c_T}{c_{T,\, \text{free}}}\,,
    \label{eq:FreeEnergy}
\end{equation}
where $c_{T,\, \text{free}} = d\, \Gamma(d/2)^2/(4\pi^d(d-1))$.
The method presented here can be tested on simple examples, and is found to produce accurate results for the free scalar field in $4d$, the $2d$ Ising model, and the $\mathrm{O} (N)$ model at large $N$ \cite{SM}. More details and illustrative examples can be found in \cite{AlessioThesis}.

\bigskip

\paragraph{Ising, XY and Heisenberg models -}
The method presented above can be used to study the $\mathrm{O}(N)$ model in $3 \leq d \leq 4$.
We consider in \eqref{eq:SumRules} the lightest scalar $\phi_i$ ($i =1,\ldots, N$) as external operator.
 We use two distinct sets of zero-temperature input: the results obtained from the $\varepsilon$-expansion \footnote{We use conformal dimensions up to order $O(\varepsilon^3)$ without resummations or Padé approximants.} and gathered in \cite{Henriksson:2022rnm}, and the results from the (zero-temperature) $3d$ bootstrap, given in \cite{El-Showk:2012cjh,Kos:2016ysd,Reehorst:2021hmp,Chester:2019ifh,Liu:2020tpf,Chester:2020iyt} for $N=1,2,3$.
To approximate the tail of heavy operators, we consider only the operators $[\phi\phi]_{n,\ell}$ in the second term of Eq. \eqref{eq:ConstraintsWithTauberian}, corresponding to the identity by channel duality.
We consider the contribution of the identity operator and of the three lightest operators in the spectrum, and one correction to the Tauberian approximation \cite{SM}.
This results in four unknowns: the three non-trivial one-point functions and the correction to the Tauberian approximation $c_1$.
All our calculations are performed by setting $m_\text{max} = 7$ in \eqref{eq:CostFunction},
which corresponds to having four constraints of the type \eqref{eq:SumRules}.
Increasing $m_\text{max}$ would result in an increased error from the Tauberian approximation, which would in turn require the inclusion of additional corrections in \eqref{eq:Tauberian}.


\begin{table}[t]
\begin{center}
\caption{OPE coefficients $a_\Delta$ of light operators in the $3d$ Ising model, compared to Monte Carlo results (MC) and previous results (PR). The value for the Tauberian correction is  $c_1 \sim -0.065$, for which the error is negligible.
}
 \renewcommand{\arraystretch}{1.25}
\begin{tabular}{  c| c || c| c| c}
\hline 
$\mathcal O$ & $\Delta_{\mathcal{O}}$ \cite{Kos:2016ysd,Reehorst:2021hmp}& This work & MC  \cite{PhysRevE.79.041142,PhysRevE.53.4414,PhysRevE.56.1642}& PR \cite{Iliesiu:2018zlz} \\  \hline 
$\epsilon$ & 1.412625(10) &0.75(15) & 0.711(3)& 0.672(74) \\ \hline 
$T_{\mu \nu}$  & 3&1.97(7) &2.092(13)  & 1.96(2)\\ \hline 
$\epsilon'$ & 3.82951(61)& 0.19(6)& 0.17(2)&0.17(2) \\ \hline 
\end{tabular}
\label{tab:3dIsingResults}
\end{center}
\end{table}

We gather our results for the $3d$ Ising model ($N=1$) in Table \ref{tab:3dIsingResults} and compare them to the Monte-Carlo values and the previous results, which relied on a different thermal bootstrap approach.All our results are consistent with previous bootstrap findings \cite{Iliesiu:2018fao}. Both our results and those of \cite{Iliesiu:2018fao} align with Monte Carlo predictions, with the sole exception of the thermal OPE coefficient of the stress-energy tensor. \footnote{After submitting our letter to the arXiv, we became aware of new, yet unpublished, high-precision Monte Carlo results for the free energy in the $\mathrm{O}(N)$ model for $1 \leq N \leq 6$ \cite{Bulgarelli:2025riv}. It would be interesting to compare these results with our predictions.}
As already observed in \cite{Iliesiu:2018fao}, the value of the stress-energy tensor contribution is close to the large $N$ approximation, where $b_T \sim - 0.459 N$ and $a_T \sim 1.923$ \cite{Sachdev:1992py} \footnote{The matching with large $N$ could also be accidental: in fact if one includes a subleading correction to the large $N$ expansion \cite{Katz:2014rla} the results seems to not match, as expected for small $N$ results.}.
The results obtained with the $\varepsilon$-expansion and the $3d$ conformal bootstrap as an input are shown in Fig. \ref{fig:Free_Energy} for the free energy density. Notice that the error estimated on the coefficient $a_d$ propagates non-trivially on the free energy; in particular it is multiplied by $N$.
We also estimated the thermal two-point function $g(\tau)$ by inputting the numerical results in the OPE: Fig. \ref{fig:2_pt} shows a comparison between the two KMS-dual channels.
The results for the OPE coefficients are presented in Fig. \ref{fig:OPE_Ising}.

Also for the XY model ($N=2$) many zero-temperature results have been obtained through the $\varepsilon$-expansion and the conformal bootstrap.
We find the following predictions for the OPE coefficients in $3d$:
%
\begin{align}
    a_{\phi^2_S} &= 0.73(14) & (&\Delta_{\phi^2_S}=1.51136(22)) \, , \\
    a_T &= 1.90(8) & (&\Delta_{T}=3) \, , \\
    a_{\phi^4_S} &= 0.20(7) & (&\Delta_{\phi^4_S}=3.794(8) )  \, .
\end{align}
The value for the Tauberian correction is  $c_1 \sim -0.0539$, for which the error is negligible.
The free-energy density can be calculated using Eq. \eqref{eq:FreeEnergy}, and the results are shown in Fig. \ref{fig:Free_Energy}. 

We performed the same calculations for the Heisenberg model ($N=3$), using the input from the $\varepsilon$-expansion and the conformal bootstrap.\
We obtain the following results for the OPE coefficients in $3d$:
%
\begin{align}
    a_{\phi^2_S} &=  0.76(14) & (&\Delta_{\phi^2_S}=1.59489(59)) \, , \\
    a_T &= 1.81(8)  & (&\Delta_{T}=3) \, , \\
    a_{\phi^4_S} &= 0.21(7) & (&\Delta_{\phi^4_S}=3.7668(100))  \, .
\end{align}
The value for the Tauberian correction is $c_1 \sim -0.0471$, for which the error is negligible.
As for the other cases, we show the free-energy density in Fig. \ref{fig:Free_Energy}.
The results for the OPE coefficients of the XY and the Heisenberg models are presented in Fig. \ref{fig:OPE_Coefficients}.
Note again that the values of $a_T$ for these models closely follow the large $N$ prediction.
The asymptotic behavior is not strictly monotonic, nonetheless the qualitative dependence of $a_T$ on $N$ aligns with the findings of \cite{Bulgarelli:2025riv}.

\bigskip

\paragraph{Discussion -}
In this letter, we propose a numerical method for computing thermal OPE coefficients, which we apply to the critical $\mathrm{O}(N)$ models for $N=1,2,3$. In particular, we extract the free-energy density of the system in $3 \leq d \leq 4$ as well as the OPE coefficients of the lightest operators.
In the case of the $3d$ Ising model ($N=1$), our results can be compared with previous studies, while for $N = 2,3$ we produce new predictions.

There are several directions to explore following this work.
The methods presented here can be applied to different models. Motivated by recent progress in the context of holographic black holes \cite{Esper:2023jeq,Dodelson:2023vrw,Bobev:2023ggk,Dodelson:2023nnr,Ceplak:2024bja},
it would be interesting to study the thermal $\mathcal{N} = 4$ super Yang-Mills and ABJM theories, for which a plethora of zero-temperature CFT data is available in the literature \cite{Gromov:2015wca,Gromov:2023hzc,Chester:2023ehi, Aharony:2008ug, Bobev:2022eus}.
Moreover, it was shown in \cite{Barrat:2024aoa} that the bootstrap problem in the presence of a temporal line defect is very similar to the one discussed in this letter.
The exploration of this direction is crucial because of low-energy applications \cite{Affleck:1995ge,Sachdev:2010uj,Sachdev:2024iux} and holographic interpretations \cite{Witten:1998zw}.
In the case of the Maldacena--Wilson line \cite{Maldacena:1998im}, a great amount of CFT data has been extracted recently \cite{Cavaglia:2021bnz,Cavaglia:2022qpg}. 
Furthermore we are currently working on improving the precision on the numerical results in tandem with developing an analytical approach which will provide the next corrections in the Tauberian approximation \cite{Barrat:2025nvu}.

The strategy of this letter could be adapted to all these configurations, which also provide a good stage for improving the precision on the numerical results.

Finally, recently many different directions to study finite temperature effects in CFTs were proposed \cite{Petkou:2018ynm,Petkou:2021zhg,Benjamin:2023qsc,Karydas:2023ufs,David:2023uya,David:2024naf,Benjamin:2024kdg,Buric:2024kxo,Alkalaev:2024jxh}. It would be interesting to compare and possibly incorporate these techniques with the method proposed in this paper.

\bigskip
\begin{acknowledgments}
\paragraph{Acknowledgments -} It is a pleasure to thank Carlos Bercini, David Berenstein, Andrea Bulgarelli, Michele Caselle, Simone Giombi, Theo Jacobson, Daniel Jafferis, Igor Klebanov, Zohar Komargodski, Juan Maldacena, Alessandro Nada, Sridip Pal, David Poland, Silviu Pufu, Leonardo Rastelli,Volker Schomerus, David Simmons-Duffin, Ning Su, Zhengdi Sun for interesting discussions and suggestions. We especially thank Simone Giombi and Igor Klebanov for pointing out \cite{HMaster} and sharing the results with us and Andrea Bulgarelli, Michele Caselle, and Alessandro Nada for sharing with us their preliminary results from \cite{Bulgarelli:2025riv}.
JB and EP's work is supported by ERC-2021-CoG - BrokenSymmetries 101044226. EM and EP's work is funded by the Deutsche
Forschungsgemeinschaft (DFG, German Research Foundation) – SFB 1624 –
“Higher structures, moduli spaces and integrability” – 506632645. JB, EM, AM and EP have benefited from the German Research Foundation DFG under Germany’s Excellence Strategy – EXC 2121 Quantum Universe – 390833306. 
AM thanks the Simons Center for Geometry and Physics, Yale University, Princeton University, Caltech, UCLA and UCSB for hospitality during the final stages of this work.
\end{acknowledgments}

\bibliography{onft}
\end{document}


\newcommand{\JB}[1]{\textcolor{orange}{[JB: #1]}}
\newcommand{\EP}[1]{\textcolor{red}{[EP: #1]}}
\newcommand{\EM}[1]{\textcolor{olive}{[EM: #1]}}
\newcommand{\AM}[1]{\textcolor{blue}{[AM: #1]}}

\newcommand{\op}[1]{\mathcal{#1}}
\newcommand{\de}{\mathrm d}
\newcommand{\Tr}{\operatorname{Tr}}
\newcommand{\ket}[1]{\left |#1 \right \rangle }

\title{Supplemental Material}
\maketitle
\section{Estimating the numerical error}

As mentioned in the main text, the numerical estimations are affected by two sources of errors.
The first one is a numerical error due to the minimization of the cost function, to which we will refer to as \textit{statistical error}.
The second one is a \textit{systematic error}, resulting from the fact that the tail of heavy operators is approximated.

The statistical error is the easiest to estimate.
For the cost function $\eta(\{\mathtt \omega_i\})$, it is clear that different sets of random weights $\{\mathtt \omega_i\}$ will produce slightly different results for the OPE coefficients.
It is therefore natural to address this error by performing the minimization procedure on different sets of random coefficients, and take the mean value as the final result with the (square root of the) variance as the statistical error. The same procedure was also recently used in the context of zero-temperature bootstrap \cite{Poland:2023bny}.

The systematic error is more intricate to estimate.
It stems from the corrective terms in the Tauberian approximation for the tail of heavy operators.
When minimizing the cost function $\eta(\{\mathtt \omega_i\})$, the results depend on three parameters:
\begin{itemize}
    \item The \textit{number of derivatives} $m_\text{max}$.
    It was shown in \cite{Marchetto:2023xap} that the contribution of the tail of heavy operators increases with the number of derivatives.
    The error can thus be minimized by choosing the smallest $m_\text{max}$ possible.
    In practice, we tune this parameter to have as many equations as unknowns;
    \item The \textit{number of corrections to the Tauberian approximation} $c_1, \ldots$.
    As mentioned in the main text, the asymptotic approximation can be corrected with terms of the form $c_i/\Delta^{\alpha_i}$, where $c_i$ and $\alpha_i$ are theory-dependent numbers.
    In the case of the $\mathrm O(N)$ model, the scaling dimension of the lightest scalar $\phi$ varies in the range $\Delta_\phi \in [1/2, 1]$.
    Since this operator is very light, we use one Tauberian correction, as the next term of the form $c_2/\Delta^{\alpha_2}$ (with $\alpha_2 > 1$) is suppressed. In the next section, the effect of the subleading Tauberian corrections is demonstrated for the $4d$ free scalar case.  A systematic, theory-independent discussion is reserved to future studies;
    \item The \textit{cut-off conformal dimension} $\Delta_\text{max}$. The parameter $\Delta_\text{max}$ controls the number of operators excluded in the Tauberian approximation.
    Selecting an appropriate $\Delta_\text{max}$ is generally challenging, as there is no \emph{a priori} method for determining its optimal value.
    In our case, we tested different values in the range $\Delta_\text{max} \in (3,10)$ and chose the one that yielded the most stable numerical minimization, i.e., the one with the least variation across different random initializations.
    To illustrate this, Tab. \ref{tab:resultsvarying} presents the results for the stress-energy tensor coefficient $a_T$ for $m_\text{max} =\{ 5,7,9,11 \}$ and for $\Delta_\text{max} =\{3,4,6\}$ (the most stable solutions). For $\Delta_\text{max} = 6$, the value of $a_{\epsilon'}$ is used as an input to avoid convergence to false minima of the cost function. Notably, increasing $m_\text{max}$ makes the procedure increasingly unstable, as the contribution of heavy operators becomes more significant. 
    The case $\Delta_{\text{max}}=3$ corresponds to considering only $\epsilon$ and $T^{\mu \nu}$ in the light sector. Note that the inclusion of $\epsilon'$, leads to significant improvements.
    In conclusion, selecting only three light operators is the optimal choice for ensuring numerical stability, measured in terms of the variance. Note that an alternative approach to estimate $a_{T}$ could involve taking a variance-weighted average over different choices of $\{\Delta_{\text{max}}, m_{\max}\}$. However, the highlighted value in Tab. \ref{tab:resultsvarying} would still dominate and the final result would be consistent with our current estimate. Finally, it should be noticed that the results available from the current state of art of the zero-temperature bootstrap naturally sets a practical upper bound to $\Delta_{\text{max}}$.

\end{itemize}

\begin{table}[b]
        \begin{center}
        \caption{OPE coefficients $a_T$ in $3d$ Ising for $m_\text{max} = \{5,7,9,11\}$ and  for $\Delta_\text{max} =\{3,4,6\}$ . The data is presented as \texttt{result(variance)}. For $\Delta_\text{max}=6$, $m_\text{max}=7$, there are more unknowns than equations and the minimization procedure fails. We highlighted the choice of parameters made in this letter.
        }
        \renewcommand{\arraystretch}{1.5}
        \begin{tabular}{|c| c| c| c| c|}
        \hline 
        \multicolumn{5}{|c|}{$a_T (\text{variance})$} \\
        \hline
        \diagbox{$\Delta_\text{max}$}{$m_\text{max}$} & 5 &  7 & 9 & 11\\  \hline
        3 & $1.492(1 \cdot 10^{-10})$ & $1.599(3 \cdot 10^{-2})$ & $1.66(6 \cdot 10^{-2})$ & $1.69(4 \cdot 10^{-2})$ \\
        \hline
        4  & no solution &$\bm{1.973(2 \cdot 10^{-18})}$ &  $2.038 (4 \cdot 10^{-4})$ &  $2.068 (2 \cdot 10^{-3})$ \\ \hline 
        6 & no solution  & no solution & $2.053(5 \cdot 10^{-4})$ & $2.105 (1 \cdot 10^{-3})$\\ \hline 
        \end{tabular}
        \label{tab:resultsvarying}
        \end{center}
    \end{table}

Once these parameters are set, we can estimate the systematic error, which is here dominant with respect to the statistical one.
There are three sources of error:
\begin{enumerate}
    \item[I.] In the heavy sector, we approximated the spectrum by modeling the heavy operators as those of a generalized free field, using the relation $\Delta = 2\Delta_\phi+2n+ \ell$. The dominant contributions in this sector come from large-spin operators, as indicated by the inversion formula \cite{Iliesiu:2018fao}. To estimate the error introduced by neglecting anomalous dimensions, we focus on these large-spin operators.
    For such operators, the approximation used correctly captures the asymptotic behavior \cite{Fitzpatrick:2012yx,vanRees:2024xkb}, but subleading corrections of order $O(1/\ell)$ must be considered. The first spin correction for the $\mathrm{O}(N)$ models studied in this work can be computed \cite{Fitzpatrick:2012yx}, allowing us to assess its impact on the coefficients $a_\Delta$ and estimate the relative error introduced by neglecting anomalous dimensions in the heavy sector.
    To illustrate this concretely, we consider the case of the stress-energy tensor OPE coefficient in the $3d$ Ising model, where the relative error due to this effect is approximately $\sim 2 \%$.
    \item[II.]
    Limiting the Tauberian approximation to a single correction introduces an error in estimating the contribution of heavy operators. To gauge the impact of the next-order correction in the Tauberian theorem on the OPE coefficient estimates, we run our numerical procedure under the assumption that the second-order coefficient, $c_2$, is of the same order as $c_1$ and that the exponent of $\Delta$ remains close to its free-field value. This assumption is justified by comparison with the inversion formula \cite{Iliesiu:2018fao}.
    To remain agnostic about the sign of $c_2$, we adopt the most conservative estimate of the error. In the case of the stress-energy tensor in the $3d$ Ising model, this leads to an estimated relative error of approximately $\sim 3\%$.
    \item[III.] As already highlighted previously, in the heavy sector we do not consider the full spectrum of operators but we focus on the dominant subset of double-twist operators, $[\phi \phi]_{n,\ell}$ (see \cite{Fitzpatrick:2012yx} for a comprehensive discussion of the heavy spectrum). These operators serve as the dual-channel counterpart of the identity operator: in the regime $\tau \sim 0$, the identity dominates, whereas for $\tau \sim \beta$, the double-twist operators take over. In the context of the $\varepsilon$-expansion, it can be shown that the contributions of other operators are suppressed in $\varepsilon$.
    For instance, in the $3d$ Ising model, we neglect operators of the form $\epsilon^m$, which, in principle, also contribute to the heavy sector. However, estimating their relevance requires knowledge of their OPE coefficients. To address this, we draw insights from the large $N$ limit, where these operators correspond to $\sigma^m$, $\sigma$ being the Hubbard-Stratonovich field. We approximate their contribution using large $N$ OPE coefficients, which become increasingly accurate for larger $N$:
    \begin{equation} a_{\sigma^m} = \frac{m_{\text{th}}^{2m}}{\Gamma(2m+1)} \ , \end{equation}
    where $m_{th}$ is the thermal mass estimated in Eq. \eqref{eq:TM}. As expected, this trajectory contributes less to the OPE coefficients of interest compared to I. and II. To be concrete, the relative error in the thermal OPE coefficient of the stress-energy tensor in the $3d$ Ising model due to this effect is approximately $\sim 0.5 \%$.
\end{enumerate}
The total error is estimated by accounting for effects I, II, and III together. To ensure a conservative estimate, we compute the squared sum of the three relative errors. This approach is cautious, as the three sources of error are not entirely independent. Notably, Tauberian corrections II already incorporate the anomalous dimensions of heavy operators, a phenomenon known as pole shifting, and predicted by the inversion formula \cite{Iliesiu:2018fao}.
A more refined analysis of the Tauberian tail and a more precise estimation of errors arising from the heavy spectrum would require more sophisticated methods. We explore this issue from different perspectives, particularly within a perturbative framework, in an upcoming work \cite{NewAnalytic}.

Errors in the zero-temperature data also propagate through the minimization procedure. In particular, the $3d$ results obtained using input from the $\varepsilon$-expansion are expected to be less precise than those derived from the $3d$ bootstrap, especially for $N=1,2,3$. The uncertainty in the $\varepsilon$-expansion zero-temperature data can be estimated by comparing it with the bootstrap results in $d=3$ \footnote{Error bars on zero-temperature bootstrap results are extremely small and negligible compared to all other sources of error considered in this work.}.
By varying the zero-temperature input data from the $\varepsilon$-expansion within its error bars, we can assess how these uncertainties propagate into the thermal OPE data. We observe that deviations in the conformal dimensions of operators are significantly smaller than the propagation of errors in the central charge, which plays a crucial role in computing the free energy across dimensions.

\section{Warm-ups: $4d$ free scalar and $2d$ Ising model}

\paragraph{$4d$ free scalar -}
In order to test the validity of the method presented in this letter, we apply it on simple models where analytical results are available.
The simplest case is the free scalar field theory in four dimensions.
The two-point function can be computed explicitly by using the method of images, and the thermal OPE data can be extracted  directly .
It is interesting to note that the solution corresponding to the free scalar field can be analytically bootstrapped, which implies unicity for the solution of the thermal bootstrap problem \cite{Marchetto:2023fcw}.
When reduced to zero spatial distance, the two-point function of fundamental scalars reads
\begin{equation}
    \langle \phi(\tau) \phi(0) \rangle_\beta = \frac{\pi^2}{\beta^2} \csc^2\left(\frac{\pi \tau}{\beta} \right) \,.
\end{equation}
Only double-twist operators appear in the OPE between the two fundamental scalars.
Furthermore, the equation of motion $\Box \phi = 0$ allows only currents to be present, i.e., operators of the (schematic) type $\phi \partial^{\mu_1}\ldots \partial^{\mu_J} \phi$. $J$ is the spin and the conformal dimensions are given by $\Delta = 2 + J$.

To obtain predictions for the OPE coefficients using the method presented in this letter, we use a different number of corrections to the Tauberian theorem.
This is a useful exercise, as increasing the number of corrections demonstrates that the only error arises from approximating the tail of heavy operators.
In free theory, the spectrum is indeed exactly given by the double-twist operators, with integer scaling dimensions.
The Tauberian asymptotic takes the form
\begin{equation}
    a_{\Delta}^{\text{heavy}} = \Delta \left(1+\frac{c_1}{\Delta}+\frac{c_2}{\Delta^2}+\frac{c_3}{\Delta^3}+\ldots \right)\,.
\end{equation} 
Recall that only the first correction is universal, but in this case further corrections can be added since the anomalous dimensions vanish.
\begin{figure*}[t]
\centering
\begin{subfigure}[t]{.50\textwidth}
   \vspace*{-5.68cm}\includegraphics[width=\textwidth]{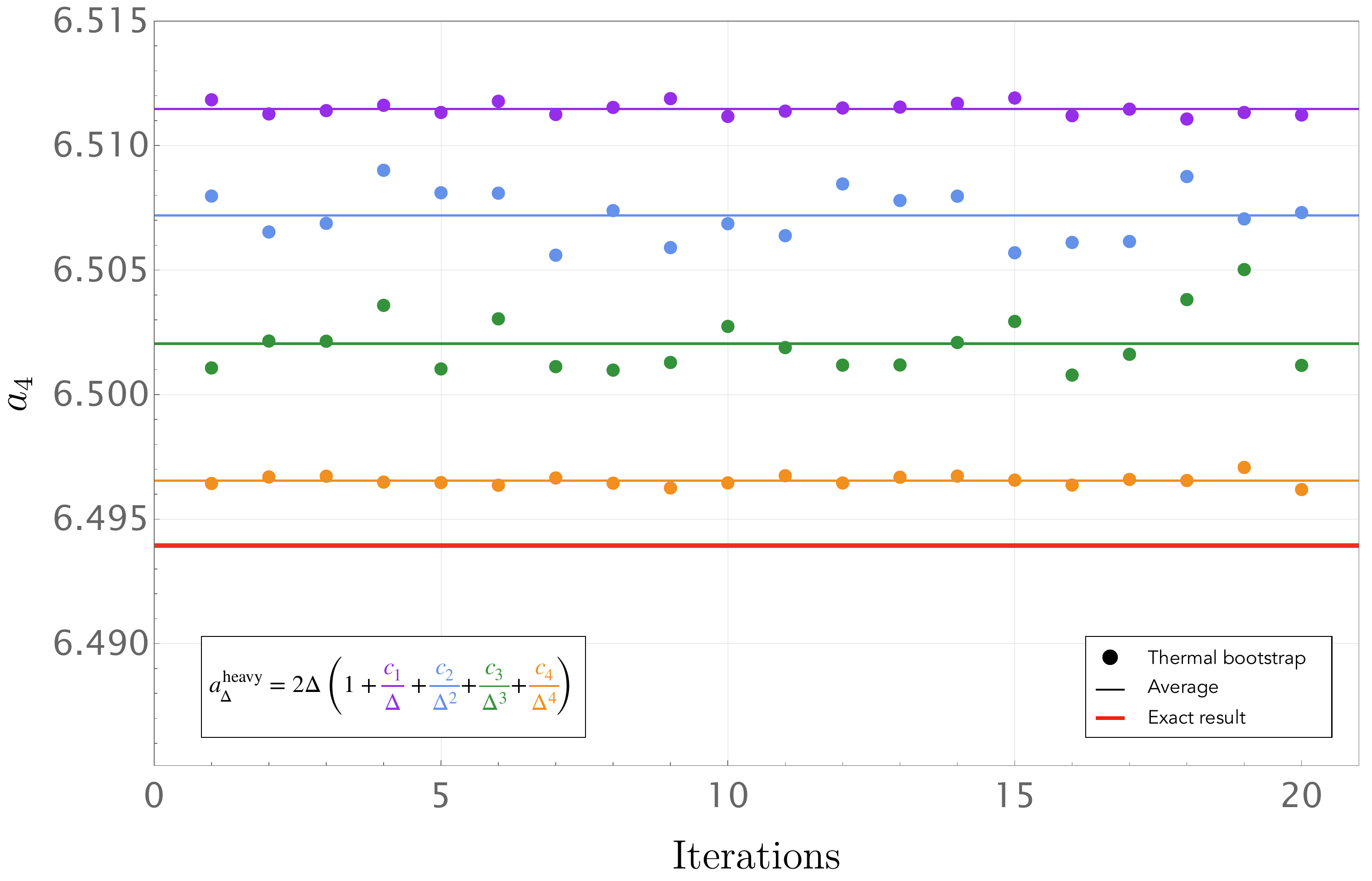}
  \caption{}\label{fig:TaubcorrFree}
\end{subfigure}%
\hfill
\begin{subfigure}[t]{.48\textwidth}
    \includegraphics[width=\textwidth]{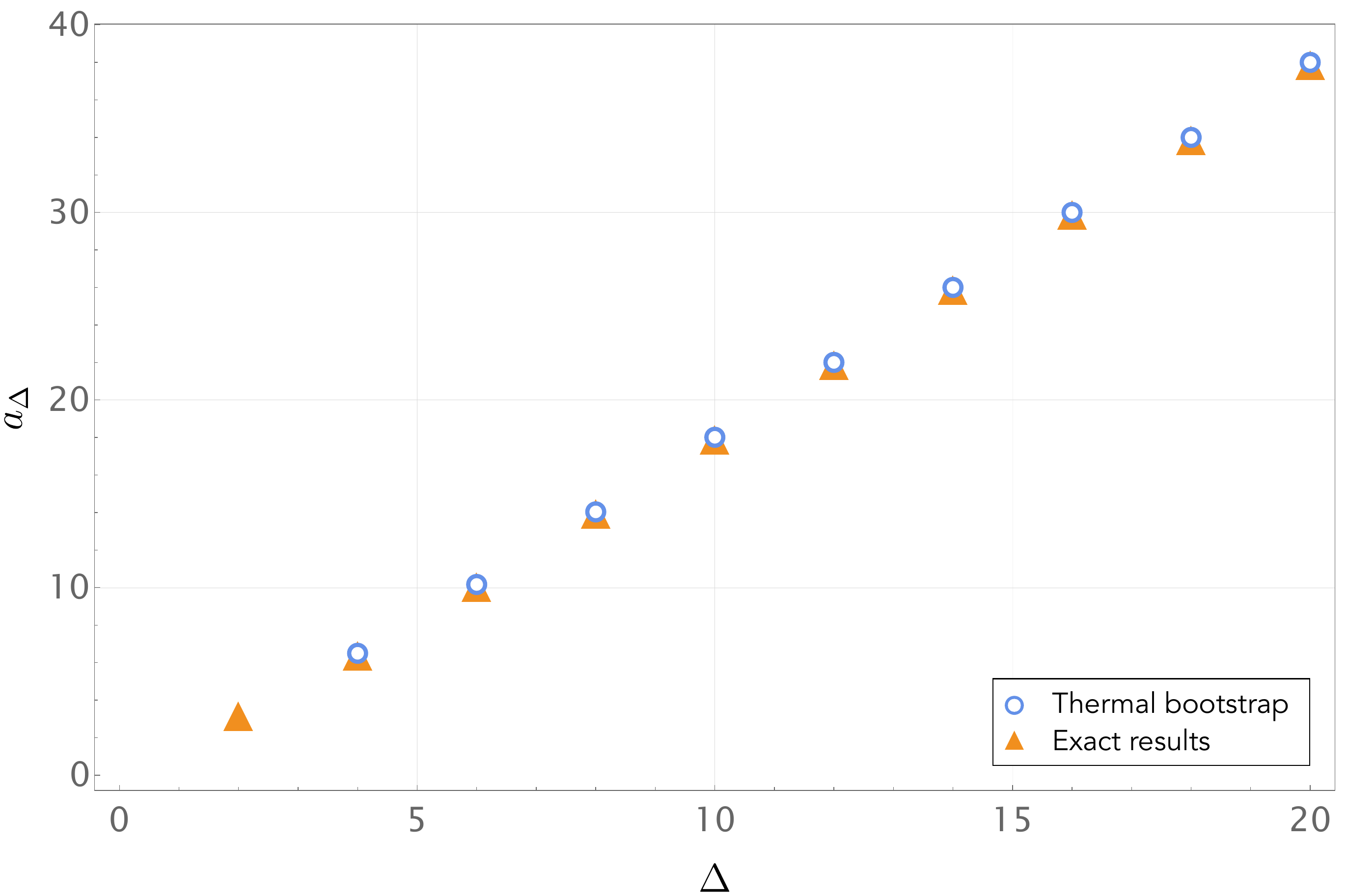}
    \caption{}\label{fig:FreeTops}
\end{subfigure}
  \caption{\emph{ \textbf{Left panel (a):} Stress-energy tensor contribution in the two-point function of fundamental scalars in the free theory for different approximations for heavy operators. \textbf{Right panel (b):} Numerical vs analytical predictions for free scalar theory. The operator $\phi^2$ ($\Delta = 2$) contributes in the two-point function as a constant and it is therefore not constrained by KMS.}}
  \label{Fig:FreeTheory}
\end{figure*}

The results of this analysis are presented in Fig. \ref{Fig:FreeTheory}, where we compare the exact results with the numerical estimations.
Notice that the discrepancy between the two decreases as the number of corrections to the Tauberian approximation increases.

\begin{figure*}[t]
\centering
\begin{subfigure}[t]{.50\textwidth}
   \vspace*{-5.58cm}\includegraphics[width=\textwidth]{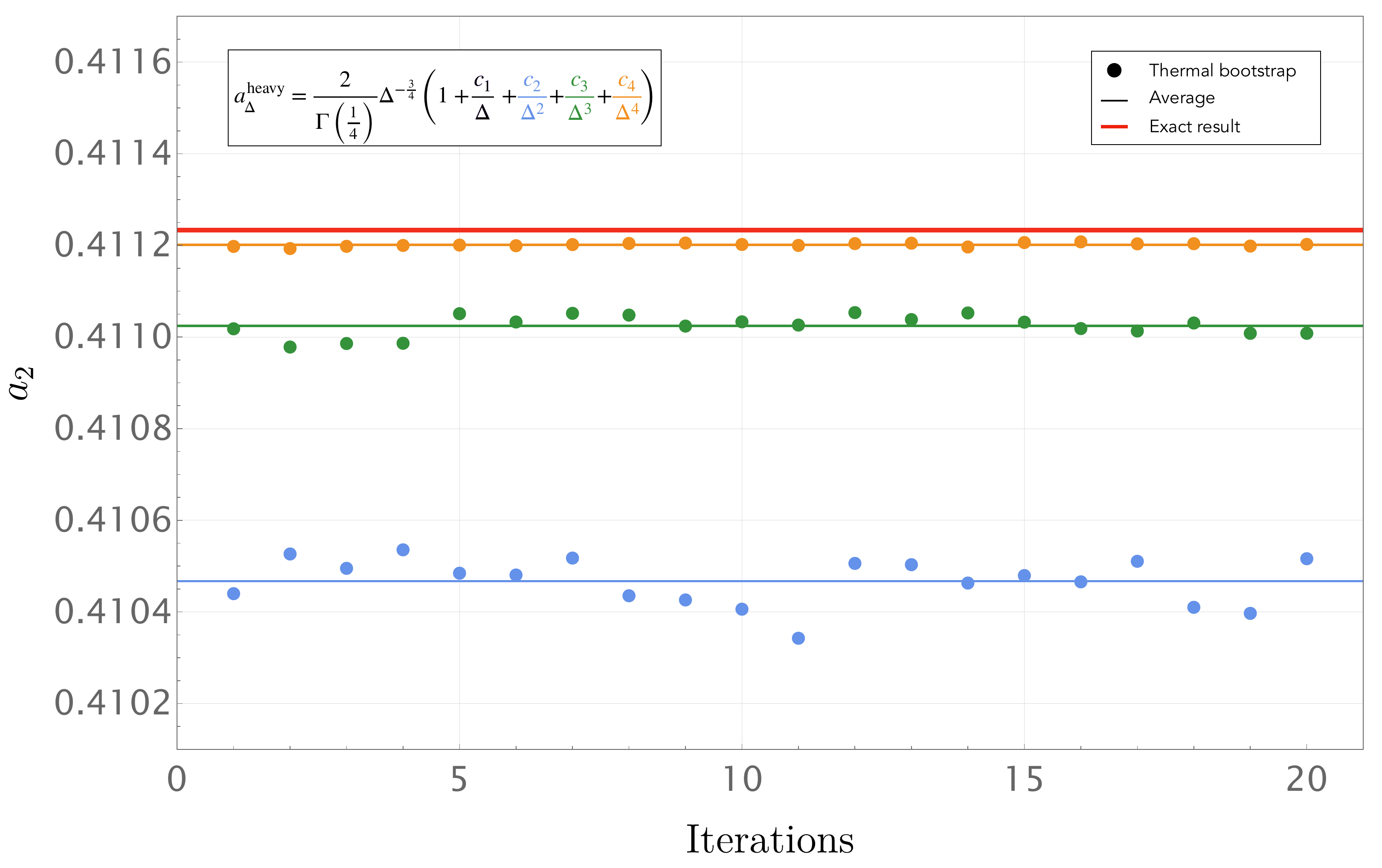}
  \caption{}\label{fig:TaubcorrFree}
\end{subfigure}%
\hfill
\begin{subfigure}[t]{.48\textwidth}
    \includegraphics[width=\textwidth]{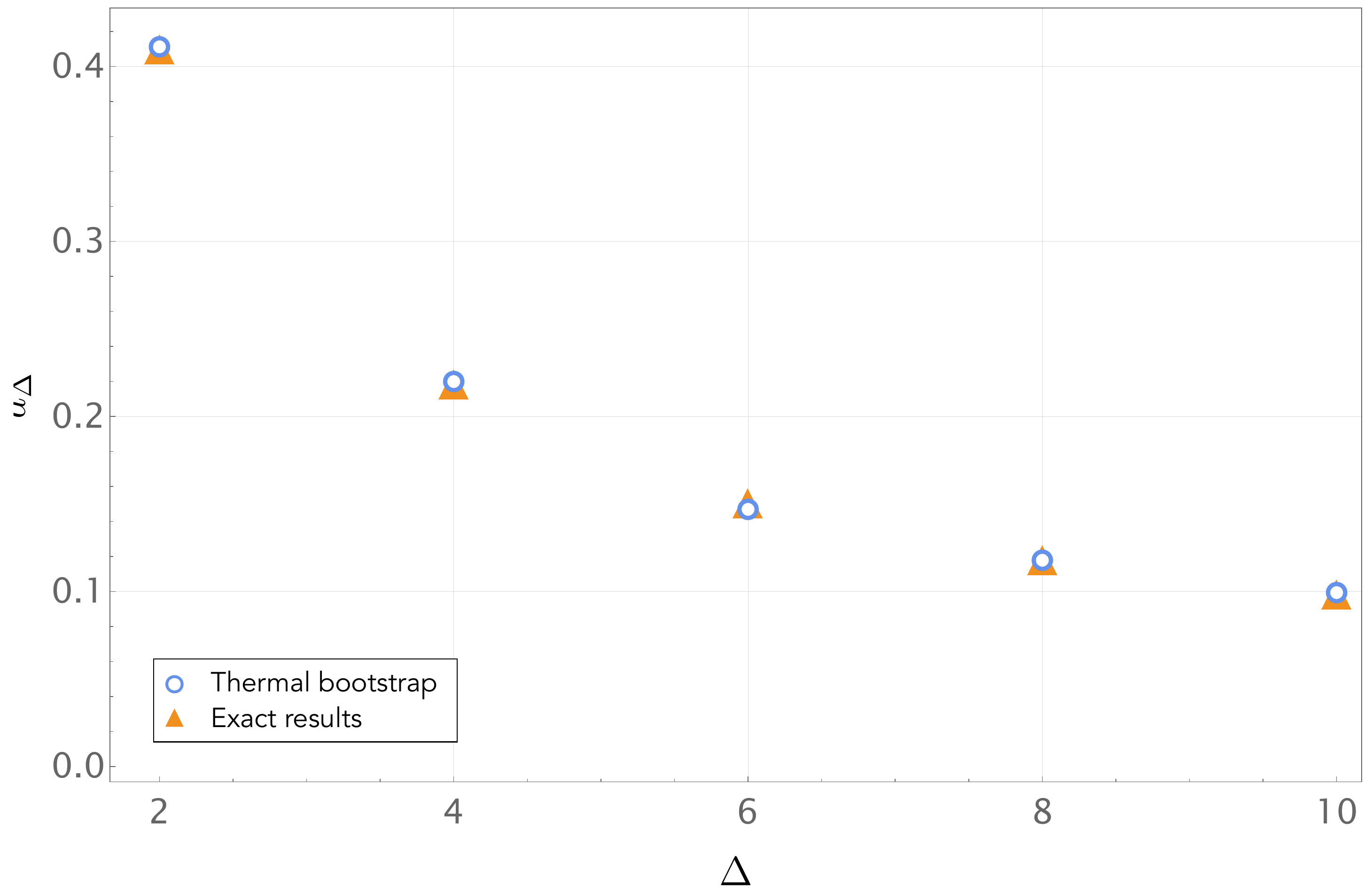}
    \caption{}\label{fig:FreeTops}
\end{subfigure}
  \caption{\emph{ \textbf{Left panel (a):} Stress-energy tensor contribution in the two-point function of lightest scalars in the $2d$ Ising model for different approximations for heavy operators. \textbf{Right panel (b):} Numerical vs analytical predictions for the $2d$ Ising model.}}
  \label{Fig:2dIsing}
\end{figure*}

\paragraph{$2d$ Ising model -}
Another useful model for testing the predictions of the numerical method is the $2d$ Ising model, where the results can be extracted analytically as well.
In this case, the one-point functions are all vanishing, except for the operators of the vacuum module.
This follows from the existence of an anomalous conformal map between the plane and the cylinder.
The two-point function of the Virasoro primary field $\sigma$, with conformal dimension $\Delta_\sigma = 1/8$, is given by
\begin{equation}
    \langle \sigma(\tau) \sigma(0) \rangle_\beta
    =
     \left|\frac{\pi}{\beta}\csc \left(\frac{\pi}{\beta} \tau\right)\right |^{1/4}\,.
\end{equation}
As mentioned above, the non-vanishing thermal one-point functions correspond to the operators $1, T^{\mu \nu},T^{\mu \nu}T^{\rho \sigma}, \ldots $ of the vacuum module.
The conformal dimensions of these operators are $0<\Delta \le J$ and $\Delta \in 2 \mathbb N$, and their one-point functions are proportional to the central charge.
The form of the corrections to the Tauberian approximation can also be predicted, since the conformal dimensions are all integer-valued.
We perform the same analysis as in the free scalar case and present the comparison between exact and numerical results in Fig. \ref{Fig:2dIsing}.

\section{Large $N$ analysis}

\begin{figure*}[t]
\centering
\begin{subfigure}[t]{.49\textwidth}
   \vspace*{-5.6cm}\includegraphics[width=\textwidth]{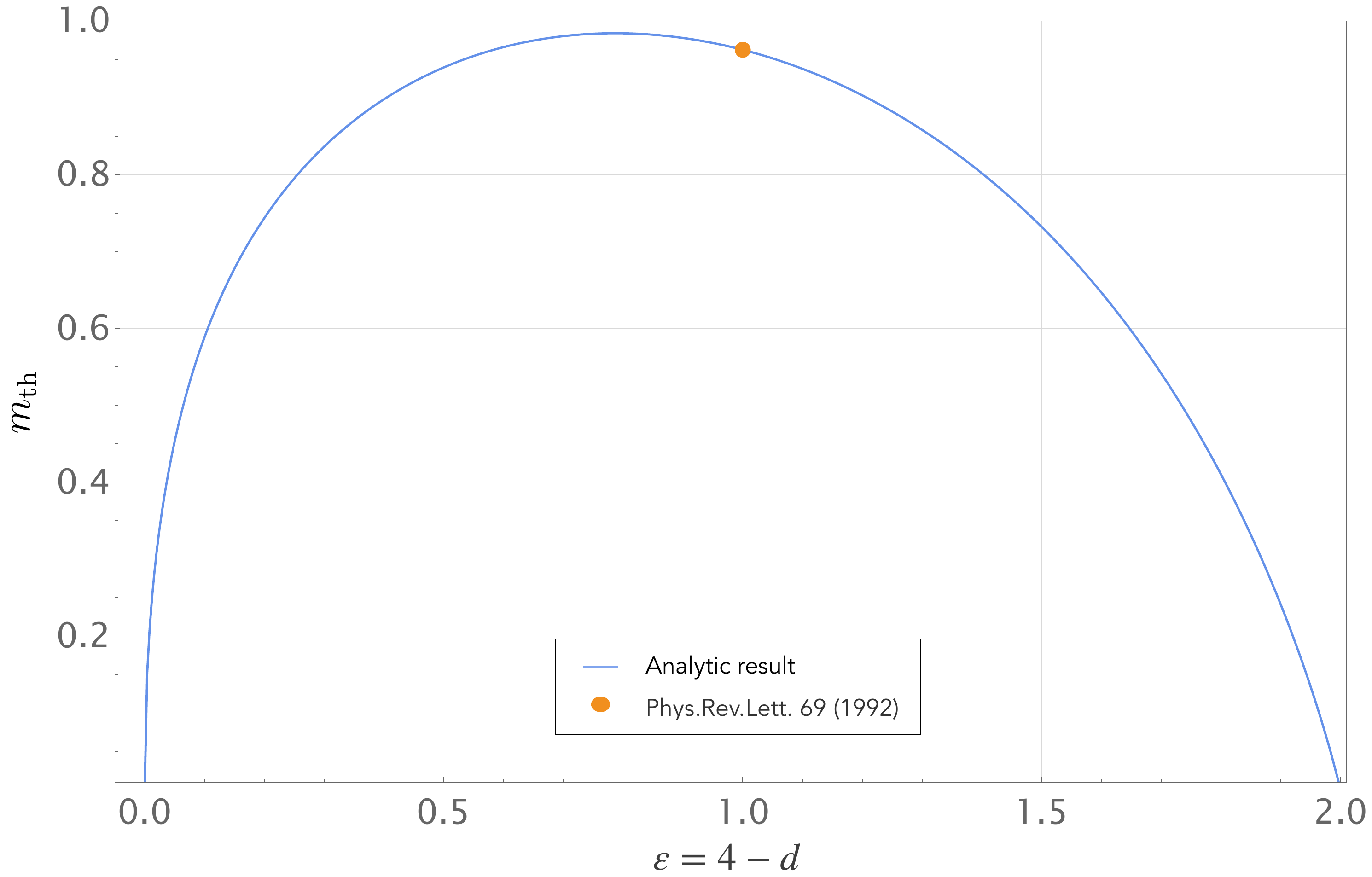}
  \caption{}\label{fig:thermalmassesLargeN}
\end{subfigure}%
\hfill
\begin{subfigure}[t]{.49\textwidth}
    \includegraphics[width=\textwidth]{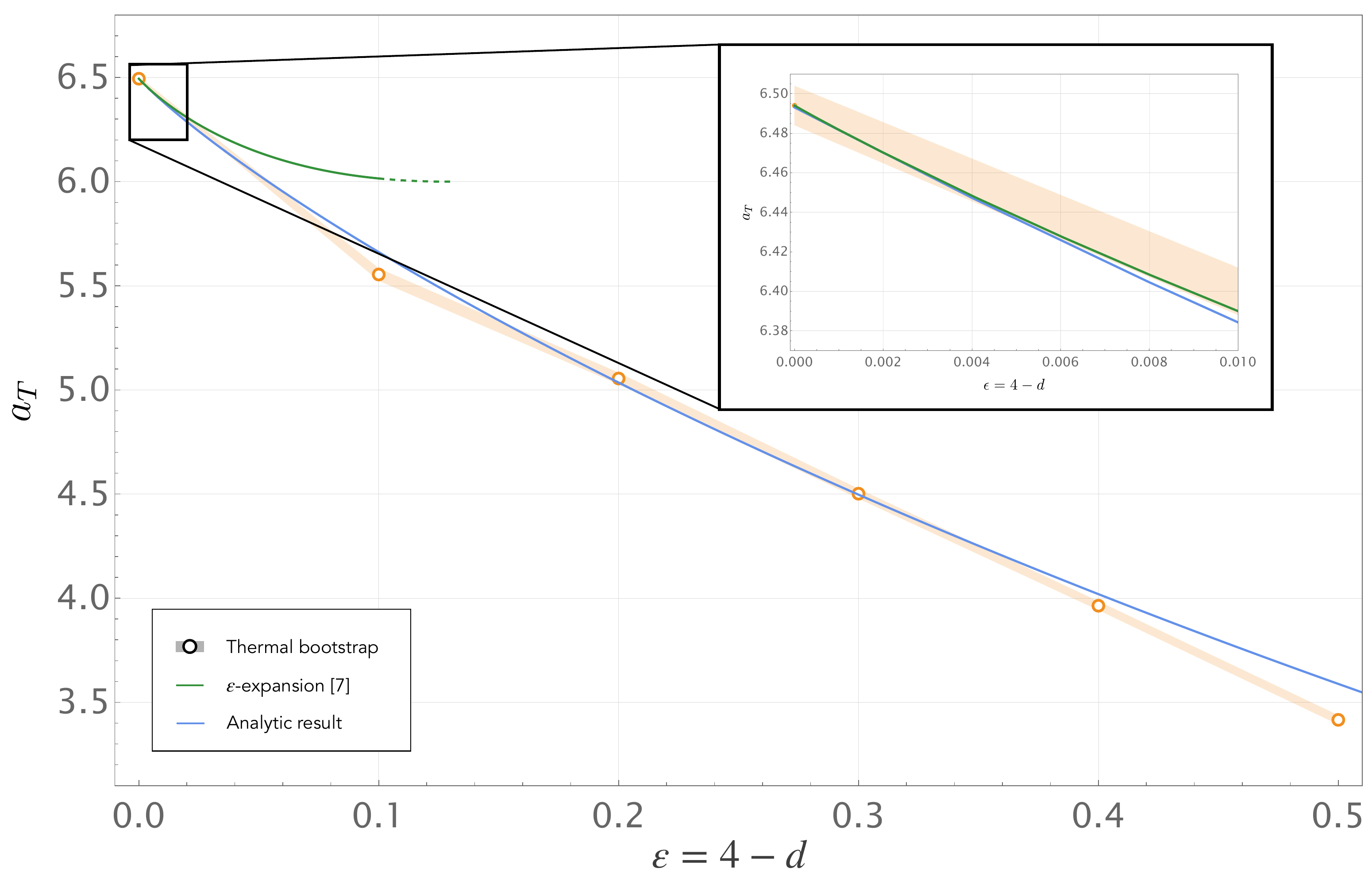}
    \caption{}\label{fig:atlargeN}
\end{subfigure}
  \caption{\emph{ \textbf{Left panel (a):} Thermal mass as a function of the spacetime dimension. \textbf{Right panel (b):} Stress-energy tensor OPE coefficient in the two-point function of fundamental scalars in the $\mathrm O(N)$ model at large $N$ as a function of the dimension. We compare numerical, analytical and $\varepsilon$-expansion results.
  }}
  \label{Fig:LargeN}
\end{figure*}

The $\mathrm O(N)$ model drastically simplifies  in the limit $N \to \infty$, where exact results can be extracted by using the Hubbard-Stratonovich formulation of the Lagrangian
\begin{equation}
    \mathcal L = \frac{1}{2}(\partial \phi_i)^2+ \frac{1}{2}\sigma \phi_i \phi_i \,.
\end{equation}
The momentum-space propagator is given by
\begin{equation}
    G_{ij}(\omega_n, \vec k) = \frac{\delta_{ij}}{\omega_n^2+\vec k^2+m_{\text{th}}^2} \,,
\end{equation}
where $m_{\text{th}}^2 = \langle \sigma \rangle_{\beta}$.
The Fourier transform can be performed and leads to a sum over free massive propagators in $d$ dimensions:
\begin{equation}\label{eq:2ptlargeN}
   \langle \phi_i(\vec x,\tau) \phi_j(0,0) \rangle_\beta  = \delta_{ij}\left(\frac{m_{\text{th}}}{2 \pi}\right)^{d-2}\sum_{m \in \mathbb Z}  \frac{ \mathrm K_{(d-2)/2}(m_{\text{th}}\sqrt{x^2+(\tau+m\beta)^2})}{ \left(m_{\text{th}}\sqrt{x^2+(\tau+m\beta)^2}\right)^{(d-2)/2}}\,.
\end{equation}
The two-point function can now be fixed by computing the thermal mass.
In order to do so, we calculate the saddle point for the partition function
\begin{equation}
    Z = \int \, D \sigma \ e^{-\frac{N}{2}\text{Tr}\log \left(\Box+\sigma\right)} \,.
\end{equation}
While the saddle point is $\sigma = 0$ on $\mathbb{R}^d$, the situation is different on $S^1_\beta \times \mathbb{R}^{d-1}$, where $\sigma$ is a non-vanishing constant.
By imposing 
\begin{equation}
    \frac{\partial}{\partial \sigma} \text{Tr}\log \left(\Box+\sigma\right) = \sum_{n = -\infty}^{\infty} \int \frac{\de^{d-1}p}{(2 \pi)^{d-1}} \frac{1}{\omega_n^2 +\vec p^2+\sigma} = 0 \,,
\end{equation}
we obtain a sum over Bessel functions. Physically we are requiring that the contribution of $\phi^2$ is absent in the two-point functions of fundamental scalars at large $N$.
This sum can be analytically performed in $3d$, and in this case the minimization of the effective action yields \cite{Sachdev:1992py,Iliesiu:2018fao,Petkou:2018ynm}
\begin{equation}\label{eq:TM}
    -m_{\text{th}}^{(3d)} = 2 \log \left(1-e^{-m_{\text{th}}^{(3d)}}\right)  \hspace{1 cm}\Rightarrow \hspace{1 cm} m_{\text{th}}^{(3d)} = 2 \log \left(\frac{1+\sqrt 5}{2}\right) \,.
\end{equation}
For $d \neq 3$, the sum can be performed numerically at a high precision, as displayed in Fig. \ref{fig:thermalmassesLargeN}.

Once the thermal mass is known, the two-point function can be expanded to extract the thermal OPE coefficients.
The OPE coefficient associated with the stress-energy tensor is presented Fig. \ref{fig:atlargeN} across dimensions.
In $3d$, it is given by \cite{Sachdev:1992py,Iliesiu:2018fao}
\begin{equation}
    a_T^{(3d)} = \frac{8}{5} \zeta(3) \,.
\end{equation}
The results can also be compared with the $\varepsilon$-expansion calculations of \cite{HMaster}, and they agree in the region $\varepsilon \ll 1$.
We compare our numerical results with the analytical ones in Fig. \ref{fig:atlargeN}. We find good agreement between the results of our thermal bootstrap procedure and the analytical result, obtained by inputting the thermal mass data plotted in Fig. \ref{fig:thermalmassesLargeN}. As expected, the analytical $\varepsilon$-expansion prediction holds close to the point $\varepsilon=0$, as shown in the zoom.  

\bibliography{sm}